\documentstyle[aps,pra,psfig,epsfig,twocolumn,floats]{revtex}
\begin{document}
\draft 

\title{Spontaneous decay in the presence of dispersing and
absorbing bodies:\\
general theory and application to a spherical cavity
}

\author{Ho Trung Dung\cite{byline}, 
Ludwig Kn\"{o}ll, and Dirk-Gunnar Welsch}
\address{
Theoretisch-Physikalisches Institut, 
Friedrich-Schiller-Universit\"{a}t Jena, 
Max-Wien-Platz 1, 07743 Jena, Germany}

\date{April 18, 2000}
\maketitle

\begin{abstract}
A formalism for studying spontaneous decay of an excited two-level 
atom in the presence of dispersing and absorbing dielectric bodies is 
developed. An integral equation, which is suitable for numerical 
solution, is derived for the atomic upper-state-probability amplitude. 
The emission pattern and the power spectrum of the emitted light 
are expressed in terms of the Green tensor of the 
dielectric-matter formation including absorption and dispersion. 
The theory is applied to the spontaneous decay of 
an excited atom at the center of a three-layered spherical cavity, 
with the cavity wall being modeled by a band-gap dielectric
of Lorentz type. Both weak coupling and strong coupling  
are studied, the latter with special emphasis on the cases 
where the atomic transition is (i) in the normal-dispersion zone near
the medium resonance and (ii) in the 
anomalous-dispersion zone associated with the band gap.
In a single-resonance approximation, conditions of the appearance of  
Rabi oscillations and closed solutions to the evolution 
of the atomic state population are derived, which are in good agreement 
with the exact numerical results.
\end{abstract}


\narrowtext

\section{Introduction}
\label{sec1}

It is well known that the spontaneous decay of an
excited atom can be strongly modified when it is placed
inside a microcavity \cite{1,2}. There are typically 
two qualitatively different regimes: the weak-coupling regime 
and the strong-coupling regime. The weak-coupling regime is 
characterized by monotonous exponential decay, the 
decay rate being enhanced or reduced compared to the 
free-space value depending on whether the atomic 
transition frequency fits a cavity resonance or not. 
The strong-coupling regime, in contrast, is characterized
by reversible Rabi oscillations where the energy of the
initially 
atom is periodically exchanged between the 
atom and the field. This usually requires that the emission is 
in resonance with a high-quality cavity mode. 
Recent progress in constructing certain types of microcavities
such as microspheres has rendered it possible to approach the 
ultimate quality level determined by intrinsic material 
losses \cite{3}, so that the question of the influence of
absorbing material on spontaneous decay has been of increasing
interest. 

Effects of material losses on the lifetime 
of an excited atom have been studied within Fermi's golden-rule 
approach \cite{4,5,6,7,7a,8,9,10}. 
In \cite{11} the mode structure of a microsphere without and with 
absorber dopant atoms, which is modeled by a constant and a Lorentzian
dielectric function respectively, is considered.   
The spontaneous emission rate and the radiation intensity 
as a function of the 
atomic transition frequency is examined in 
\cite{12} for an atom in a Fabry-Perot 
cavity filled with a Lorentz-type dielectric 
in the case of strong medium-cavity interaction  
but weak atom-field interaction.

In this paper we present a theory of the spontaneous decay
of an excited two-level atom in the presence of arbitrary
dispersing and absorbing dielectric bodies. We apply
the theory to the spontaneous decay of an atom in a 
spherical microcavity of given complex-valued refractive-index 
profile, as it is typically the case in experimental 
implementations. The formalism enables us to include in 
the theory absorption and dispersion in a consistent way 
and to give a unified treatment of spontaneous 
emission, without restriction to a particular
coupling regime.

The plan of the paper is as follows.
In Section \ref{sec2}, a recently developed quantization
scheme for the electromagnetic field in the presence of 
dispersing and absorbing dielectric bodies \cite{13,13a,14,15}
is extended in order to include in the theory the resonant
interaction of the field with a two-level atom. From
the Hamiltonian of the composed system, an integral equation
governing the temporal evolution of the upper-level-probability
amplitude of the atom is derived, the integral kernel being 
determined by the Green tensor of the classical, 
phenomenological Maxwell equations for 
the dielectric-assisted electromagnetic field. 
General expressions for the emission pattern and the 
power spectrum are derived in terms of the 
atomic parameters and, via the Green tensor, 
the cavity parameters of the dielectric-matter 
configuration. 

In Section \ref{sec3}, the theory is used to examine the 
spontaneous decay of an excited two-level atom inside a 
spherical cavity, with special emphasis on the intrinsic 
dispersion and absorption of the wall material. 
Spherical microcavities have been very attractive systems  
for both fundamental research in cavity quantum electrodynamics 
and applications in optoelectronics (see, e.g., \cite{3,26,27,28} 
and references therein). Changes in the level shifts
and lifetimes of atoms inside or near the surfaces of spherical
micro-structures have been studied theoretically \cite{29}, 
the latter also 
experimentally \cite{30}, and results on the strong-coupling
regime have been reported \cite{31,32,33}. 
The theoretical results have been  typically based on standard mode
expansion, which fails for intrinsically dispersing and absorbing   
material. 
Here, we consider a spherical three-layered structure, where
the  middle layer is assumed to be a (single-resonance) band-gap 
dielectric of Lorentz type, whereas the outer and inner layers   
are vacuum. The strength of the atom-field coupling, 
which is essentially determined by the imaginary part of the
Green tensor at the position of the atom, is 
analyzed and the positions, heights, and widths of the
possible cavity resonances are calculated. In particular, the 
most pronounced resonances are observed within the band gap,
the widths of which are proportional to the intrinsic
damping constant of the medium. In a single-resonance
approximation, conditions of the strong-coupling regime
and closed expressions for the atomic 
upper-state-population probability are derived, which
are in good agreement with the numerical solutions. 
Finally, a summary and conclusions are given in 
Section \ref{sec4}.

\section{General formalism}
\label{sec2}

\subsection{Quantization scheme}
\label{sec2.1}

Let us first consider the electromagnetic field in the
presence of dispersing and absorbing dielectric bodies
without additional atomic sources. Following \cite{13,13a,14,15}, we
represent the electric-field operator $\hat{\bf E}$ in the form
\begin{eqnarray}
\label{e10}
&\hat{\bf E}({\bf r}) = \hat{\bf E}^{(+)}({\bf r})
   + \hat{\bf E}^{(-)}({\bf r}),
\quad
\hat{\bf E}^{(-)}({\bf r})
   = \big[\hat{\bf E}^{(+)}({\bf r})\big]^\dagger&
\\[.5ex]
\label{e11}
&\displaystyle\hat{\bf E}^{(+)}({\bf r}) 
   = \int_{0}^{\infty} d\omega\,
   \underline{\hat{\bf E}}({\bf r},\omega) ,&
\end{eqnarray}
and the magnetic-field operator $\hat{\bf B}$ accordingly. 
The operators $\underline{\hat{\bf E}}$ and $\underline{\hat{\bf B}}$
then satisfy the Maxwell equations 
\begin{eqnarray} 
\label{e1}
      & \displaystyle
      \bbox{\nabla} \cdot \underline{\hat{\bf B}}({\bf r},\omega) = 0, & 
\\[.5ex]       
\label{e2}
      & \displaystyle 
      \bbox{\nabla} \cdot \left[ \epsilon_0 \epsilon({\bf r},\omega)
       \underline{\hat{\bf E}}({\bf r},\omega)\right] =
       \underline{\hat{\rho}}({\bf r},\omega) , & 
\\[.5ex]
\label{e3}
      & \displaystyle 
      \bbox{\nabla} \times \underline{\hat{\bf E}}({\bf r},\omega) = i\omega
       \underline{\hat{\bf B}}({\bf r},\omega) , &
\\[.5ex]
\label{e4}
      & \displaystyle 
      \bbox{\nabla} \times \underline{\hat{\bf B}}({\bf r},\omega) =
       -i\frac{\omega}{c^2}\,\epsilon({\bf r},\omega) 
       \underline{\hat{\bf E}}({\bf r},\omega) 
       +\mu_0 \underline{\hat{\bf j}}({\bf r},\omega), &
\end{eqnarray}
where the complex permittivity $\epsilon({\bf r},\omega)$ 
is a function of frequency and space, the real part ($\epsilon_{\rm R}$)
and the imaginary part ($\epsilon_{\rm I}$) of which satisfy 
(for any ${\bf r}$) the Kramers--Kronig relations. 
The  operator noise charge and current densities  
$\underline{\hat{\rho}}({\bf r},\omega)$ and 
$\underline{\hat{\bf j}}({\bf r},\omega)$ respectively, which are 
associated with absorption, are related to the operator noise 
polarization $\underline{\hat{\bf P}}({\bf r},\omega)$ as
\begin{eqnarray}
\label{e4.1}
        &\displaystyle
        \underline{\hat{\rho}}({\bf r},\omega) = 
         - \bbox{\nabla} \cdot 
         \underline{\hat{\bf P}}({\bf r},\omega) ,&
\\[.5ex]
\label{e4.2}
         &\displaystyle
         \underline{\hat{\bf j}}({\bf r},\omega) =
         -i\omega \underline{\hat{\bf P}}({\bf r},\omega) ,&
\end{eqnarray}
where 
\begin{equation} 
\label{e5}
       \underline{\hat{\bf P}}({\bf r},\omega) 
       = i \sqrt{\frac{\hbar
       \epsilon_0}{\pi} \epsilon_{\rm I}({\bf r},\omega)} 
       \,\hat{\bf f}({\bf r},\omega)
\end{equation}
with
\begin{eqnarray}
\label{e6}
      &\displaystyle 
      \left[ \hat{f}_i({\bf r},\omega),
       \hat{f}^{\dagger}_j({\bf r}',\omega')\right]           
       = \delta_{ij}
       \delta({\bf r}-{\bf r}') \delta(\omega-\omega') , & 
\\
\label{e7}
      &\displaystyle 
      \left[ \hat{f}_i({\bf r},\omega),
       \hat{f}_j({\bf r}',\omega')\right] 
       = 0 =
       \left[ \hat{f}^{\dagger}_i({\bf r},\omega),
       \hat{f}^{\dagger}_j({\bf r}',\omega')
       \right] . &
\end{eqnarray}
   
{F}rom Eqs.~(\ref{e1}) -- (\ref{e5}) it follows that
$\underline{\hat{\bf E}}$ can be written in the form
\begin{eqnarray}
\label{e8}
\lefteqn{
\hat{\underline{\bf E}}({\bf r},\omega)
   = i \sqrt{\frac{\hbar}{\pi\epsilon_0}}\,\frac{\omega^2}{c^2}
}
\nonumber\\[.5ex]&&\hspace{5ex}\times
   \int d^3{\bf r}'\,\sqrt{\epsilon_{\rm I}({\bf r}',\omega)}
   \,\bbox{G}({\bf r},{\bf r}',\omega)   
   \cdot\hat{\bf f}({\bf r}',\omega),  
\end{eqnarray}
and $\hat{\underline{\bf B}}$ $\!=$
$\!(i\omega)^{-1} \bbox{\nabla}\times\hat{\underline{\bf E}}$ accordingly,
where
$\bbox{G}({\bf r},{\bf r}',\omega)$
is the classical Green tensor satisfying the equation
\begin{equation}
\label{e9}
\left[\frac{\omega^2}{c^2}\,\epsilon({\bf r},\omega)
   - \bbox{\nabla}\times\bbox{\nabla}\times\right]
   \bbox{G}({\bf r},{\bf r}',\omega)
= - \bbox{\delta}({\bf r}-{\bf r}')   
\end{equation}
together with the boundary condition at infinity
[$\bbox{\delta}({\bf r})$ is the dyadic
$\delta$-function].
In this way, the electric- and magnetic-field strengths are 
expressed in terms of a continuum set of bosonic fields $\hat{\bf f}$ 
and $\hat{\bf f}^\dagger$, which play the role of the 
fundamental (dynamical) variables of the composed system
(electromagnetic field and the medium including the dissipative system)
whose Hamiltonian is
\begin{equation}
\label{e9a}
\hat{H} =  \int d^3{\bf r} \int_0^\infty d\omega \,\hbar\omega
   \,\hat{\bf f}^\dagger({\bf r},\omega)\cdot\hat{\bf f}({\bf r},\omega).
\end{equation}  
Using Eqs.~(\ref{e8}) 
[together with Eqs.~(\ref{e10}) and (\ref{e11})], we can also 
express the scalar potential $\hat{\varphi}$ and the vector potential
$\hat{\bf A}$ of the electromagnetic field in terms of the 
fundamental bosonic fields. In particular,
in the Coulomb gauge we obtain
\begin{eqnarray} 
\label{e12.1}
         &\displaystyle 
         -\bbox{\nabla} \hat{\varphi}({\bf r}) 
         = \hat{\bf E}^\parallel({\bf r}) , &
\\[.5ex]
\label{e12}
         &\displaystyle 
         \hat{\bf A}({\bf r}) = 
         \int_0^\infty d \omega \, (i\omega)^{-1} 
         \underline{\hat{\bf E}}^\perp({\bf r},\omega) 
         + {\rm H.c.}\,, &
\end{eqnarray}
where
\begin{equation}
\label{e12.2}
\hat{\bf E}^{\perp(\parallel)}({\bf r})
= \int d^3{\bf r}' \, \mbox{\boldmath $\delta$}
                         ^{\perp(\parallel)}({\bf r}-{\bf r}')
         \cdot \hat{\bf E}({\bf r}'),
\end{equation}
with $\mbox{\boldmath $\delta$}^\perp({\bf r})$ 
and 
$\mbox{\boldmath $\delta$}^\parallel({\bf r})$ being
the transverse and longitudinal $\delta$-functions respectively.

We now consider the interaction of the medium-as\-sis\-ted
electromagnetic field with additional point charges $q_\alpha$.
Applying the minimal-coupling scheme, we may write the
complete Hamiltonian in the form
\begin{eqnarray}
\label{e12.5}
\lefteqn{
\hat{H} = \int d^3{\bf r} \int_0^\infty d\omega \,\hbar\omega
   \,\hat{\bf f}^\dagger({\bf r},\omega)\cdot\hat{\bf f}({\bf r},\omega)
}
\nonumber\\&&\hspace{5ex}   
   + \sum_\alpha {1\over 2m_\alpha} \left[ \hat{\bf p}_\alpha 
   - q_\alpha \hat{\bf A}({\bf r}_\alpha) \right]
   \cdot \left[\hat{\bf p}_\alpha 
   - q_\alpha \hat{\bf A}({\bf r}_\alpha) \right]
\nonumber\\&&\hspace{5ex}
   +{1\over 2} \,\int d^3{\bf r}\, 
   \hat{\rho}_{\rm A}({\bf r}) \hat{\varphi}_{\rm A}({\bf r})
   + \int d^3{\bf r}\, 
   \hat{\rho}_{\rm A}({\bf r}) \hat{\varphi}({\bf r}) ,
\end{eqnarray}
where $\hat{\bf r}_\alpha$ is the position operator
and $\hat{\bf p}_\alpha$ is the canonical momentum operator
of the $\alpha$th charged particle of mass $m_\alpha$.
The Hamiltonian (\ref{e12.5}) consists of four terms. The first 
term is the energy (\ref{e9a}) observed when the
particles are absent. The second term is the 
kinetic energy of the particles, and the third term is 
their Coulomb energy, where the potential $\hat{\varphi}_{\rm A}$ 
can be given by
\begin{equation}
\label{e12.4}
         \hat{\varphi}_{\rm A}({\bf r}) = 
         \int d{\bf r}' \,\frac {\hat{\rho}_{\rm A}({\bf r}')}
         {4\pi\epsilon_0|{\bf r}-{\bf r}'|} \,,
\end{equation}
with
\begin{equation}
\label{e12.3}
         \hat{\rho}_{\rm A}({\bf r}) = 
         \sum_\alpha q_\alpha 
         \delta({\bf r}-\hat{\bf r}_\alpha) 
\end{equation}
being the charge density. The last term is the Coulomb energy 
of interaction of the particles with the medium. Note
that all terms are expressed in terms of the
dynamical variables $\hat{\bf f}({\bf r},\omega),
\hat{\bf f}^\dagger({\bf r},\omega),\hat{\bf r}_\alpha,
\hat{\bf p}_\alpha$.


\subsection{Dynamics of an excited two-level atom}
\label{sec2.2}

Let us consider a neutral two-level atom (position ${\bf r}_{\rm A}$,
transition frequency $\omega_{\rm A}$)
that resonantly interacts with radiation via an electric-dipole 
transition (dipole moment $\bbox{\mu}$). 
In this case, the electric-dipole approximation and the rotating
wave approximation apply, and the minimal-coupling Hamiltonian  
(\ref{e12.5}) simplifies to (Appendix A)
\begin{eqnarray}
\label{e14}
\lefteqn{
          \hat{H} = \int d^3{\bf r} \int_0^\infty d\omega \,\hbar\omega
          \,\hat{\bf f}^\dagger({\bf r},\omega)\cdot\hat{\bf f}({\bf r},\omega)
}
\nonumber\\&&\hspace{5ex}
          + \, {\textstyle{1\over 2}}\hbar\omega_{\rm A} \hat{\sigma}_z
          -\left[
          \hat{\sigma}^\dagger
          \hat{\bf E}^{(+)}({\bf r}_{\rm A})\cdot\bbox{\mu}\,  
          + {\rm H.c.}\right], 
\end{eqnarray}
where $\hat{\sigma}$, $\hat{\sigma}^\dagger$, and $\hat{\sigma}_z$ are 
the Pauli operators of the two-level atom. 

When the atom is initially in the upper state and the
rest of the system is in the vacuum, then the system wave 
function at time $t$ can be written as
\begin{eqnarray}
\label{e15}
\lefteqn{
          |\psi(t)\rangle = C_{u}(t) 
          e^{-i(\omega_{\rm A}/2)t}
          |u\rangle |\{0\}\rangle
}
\nonumber\\&&
          +\!\int\! d^3{\bf r} \int_0^\infty\!\!\! d\omega\, 
          C_{li}({\bf r},\omega,t)
          e^{-i (\omega-\omega_{\rm A}/2)t}
          |l\rangle |\{1_i({\bf r},\omega)\} \rangle,
\end{eqnarray}
where $|u\rangle$ and $|l\rangle$ respectively are the upper 
and lower atomic states, $|\{0\}\rangle$ is the 
vacuum state of the rest of the system, and $|\{1_i({\bf r},\omega)\}\rangle$ 
is the state, where the latter is excited in a single-quantum Fock state.
Here and in the following we adopt the convention of summation over repeated
vector-component indices. The Schr\"{o}dinger equation yields
\begin{eqnarray}
\label{e16}
\lefteqn{
          \dot{C}_u(t) =
          -\frac{\mu_j}{\sqrt{\pi\epsilon_0\hbar}}           
          \int_0^\infty \!\!d\omega \,\frac{\omega^2}{c^2} 
          \int d^3{\bf r}\,\Big[
          \sqrt{\epsilon_{\rm I}({\bf r},\omega)}
}
\nonumber \\&&\hspace{10ex} \times\,
          G_{ji}({\bf r}_{\rm A},{\bf r},\omega) \,
          C_{li}({\bf r},\omega,t) e^{-i(\omega-\omega_{\rm A})t}\Big] ,
\end{eqnarray}
\begin{eqnarray}
\label{e17}
\lefteqn{          
          \dot{C}_{li}({\bf r},\omega,t) =
          \frac{\mu_j}{\sqrt{\pi\epsilon_0\hbar}}\,
          \frac{\omega^2}{c^2}\,\sqrt{\epsilon_{\rm I}({\bf r},\omega)}
}
\nonumber \\&&\hspace{14ex} \times\,          
          G_{ji}^\ast({\bf r}_{\rm A},{\bf r},\omega)\,
          C_{u}(t) e^{i(\omega-\omega_{\rm A})t} . 
\end{eqnarray}
We now substitute the result of formal integration of Eq.~(\ref{e17})
[$C_{li}({\bf r},\omega,0)$ $\!=$ $\!0$] into Eq.~(\ref{e16}).
Making use of the relationship 
\begin{eqnarray}
\label{e17a}
\lefteqn{
        {\rm Im}\,G_{kl}({\bf r},{\bf r'},\omega) =
}\nonumber\\&&\hspace{5ex}
        \int d^3{\bf s}\;
        \frac{\omega^2}{c^2}\, \epsilon_{\rm I}({\bf s},\omega)
        G_{km}({\bf r},{\bf s},\omega) 
        G^\ast_{lm}({\bf r'},{\bf s},\omega), 
\end{eqnarray}
we obtain the integro-differential equation
\begin{equation}
\label{e18}
        \dot{C}_u(t) =\int_0^t dt'\, K(t-t') C_{u}(t'),
\end{equation}
with the kernel function
\begin{eqnarray}
\label{e19}
\lefteqn{
        K(t-t') =
        - \frac{k_{\rm A}^2 \mu_i \mu_j }{\hbar\pi\epsilon_0}
}
\nonumber\\&&\hspace{5ex}\times         
        \int_0^\infty d\omega \,
        e^{-i(\omega-\omega_{\rm A})(t-t')} 
        {\rm Im}\,G_{ij}({\bf r}_{\rm A},{\bf r}_{\rm A},\omega)
\end{eqnarray}
($k_{\rm A}$ $\!=$ $\!\omega_{\rm A}/c$). In the
spirit of the rotating wave approximation used
we have set $\omega$ $\!=$ $\omega_{\rm A}\!$
in the integral in Eq.~(\ref{e19}). 

Taking the time integral of both sides of Eq.~(\ref{e18}), 
we easily derive, on changing the order of integrations
on the right-hand side,
\begin{equation}
\label{e19.1}
        C_{u}(t) =\int_0^t dt'\, \bar{K}(t-t') C_{u}(t') + 1
\end{equation}
[$C_{u}(0)$ $=$ $\!1$], where 
\begin{eqnarray}
\label{e19.2}
\lefteqn{
        \bar{K}(t-t') =
        \frac{k_{\rm A}^2 \mu_i \mu_j}{\hbar\pi\epsilon_0}
}
\nonumber\\&&\hspace{2ex}\times         
         \int_0^\infty \!\!d\omega\,
        \frac{{\rm Im}\,G_{ij}({\bf r}_{\rm A},{\bf r}_{\rm A},\omega)}
        {i(\omega-\omega_{\rm A})}
        \left[e^{-i(\omega-\omega_{\rm A})(t-t')}-1 \right].
\end{eqnarray}
The integral equation (\ref{e19.1}) is a well-known Volterra equation 
of the second kind. An algorithm for solving such an 
integral equation numerically can be found, e.g., in \cite{21}. 
It is worth noting that the
integro-differential equation (\ref{e18}) and the equivalent
integral equation (\ref{e19.1}) apply to 
the spontaneous decay of an atom in the presence of an 
arbitrary configuration of dispersing and absorbing dielectric 
bodies. All the matter parameters that are relevant for  
the atomic evolution are contained, via the 
Green tensor, in the kernel functions (\ref{e19})
and (\ref{e19.2}). In particular when absorption is
disregarded and the permittivity is regarded as being 
a real frequency-independent quantity (which of course
can change with space), then the formalism yields the results 
of standard mode decomposition (see, e.g. \cite{18,19,20}).

When the Markov approximation applies, i.e., when  
in a coarse-grained description of the atomic motion
memory effects are disregarded, then we may let
\begin{equation}
\label{e.19.2a}
\frac{e^{i(\omega_{\rm A}-\omega)(t-t')}-1}
{i(\omega_{\rm A}-\omega)} \to \zeta(\omega_{\rm A}-\omega)
\end{equation}
in Eq.~(\ref{e19.2}) [$\zeta(x)$ $\!=$ $\!\pi\delta(x)$ $\!+$ 
$\!i{\cal P}/x$; ${\cal P}$ denotes the principal value], and thus
\begin{equation}
\label{e.19.2b}
\bar{K}(t-t') = - {\textstyle\frac{1}{2}} 
   \left(A - i \delta\omega\right),
\end{equation}
where
\begin{equation}
\label{e21}
         A = {2k_{\rm A}^2\mu_i\mu_j\over \hbar\epsilon_0}\, 
         {\rm Im}\,G_{ij}({\bf r}_{\rm A},{\bf r}_{\rm A},\omega_{\rm A}) 
\end{equation}
and 
\begin{equation}
\label{e21a}
\delta\omega = {2k_{\rm A}^2\mu_i\mu_j\over \pi\hbar\epsilon_0}\,
   {\cal P}\!\int_0^\infty d\omega\, 
   \frac{{\rm Im}\,G_{ij}({\bf r}_{\rm A},{\bf r}_{\rm A},\omega)}
   {\omega-\omega_{\rm A}}\,. 
\end{equation}
Substituting into Eq.~(\ref{e19.1}) for the kernel function
the expression (\ref{e.19.2b}), we obtain the familiar 
result that
\begin{equation}
\label{e20}
C_{u}(t) = \exp\!\left[-{\textstyle\frac{1}{2}}
   (A-i\delta\omega)t\right].
\end{equation}
Obviously, this result is also obtained if in the integral in 
Eq.~(\ref{e18}) $C_{u}(t')$ is replaced by $C_{u}(t)$ and then
the integral is approximated by $\zeta(\omega_A$ $\!-$ $\!\omega)$.
Note that the expressions (\ref{e21}) and (\ref{e21a}) for
the decay rate and the line shift, respectively, are
in full agreement with the results in \cite{9}. 

It is well known that the Markov approximation is
an excellent approximation for describing the radiative 
decay of an excited atom in free space. In order to study the
case where the atom is surrounded by dielectric matter, 
we assume that the atom is localized in a more or less 
small free-space region, so that the Green tensor at 
the position of the atom reads \cite{22}
\begin{equation}
\label{e19.3}
        \bbox{G}({\bf r}_{\rm A},{\bf r}_{\rm A},\omega)
        = \bbox{G}^{\rm V}({\bf r}_{\rm A},{\bf r}_{\rm A},\omega)
        + \bbox{G}^{\rm R}({\bf r}_{\rm A},{\bf r}_{\rm A},\omega) ,
\end{equation}
where 
$\bbox{G}^{\rm V}$ is the vacuum Green tensor, with
\begin{equation}
\label{e19.4}
         {\rm Im}\,\bbox{G}^{\rm V}({\bf r}_{\rm A},{\bf r}_{\rm A},\omega)
         = {\omega \over 6\pi c} \, \bbox{I} 
\end{equation}
(for the vacuum Green tensor, see, e.g., \cite{4,22b}),
and 
$\bbox{G}^{\rm R}$
describes the effects of reflections at the 
(surfaces of discontinuity of the) surrounding medium.
The contribution of
$\bbox{G}^{\rm V}$ to $\bar{K}$
can then be treated in the Markov approximation.
Application of Eqs.~(\ref{e.19.2b}) -- (\ref{e21a})
yields the well-known vacuum decay rate
\begin{equation}
\label{e19.6}
        A_0= {k_{\rm A}^3\mu^2 \over 3\hbar\pi\epsilon_0} 
\end{equation}
and a divergent contribution to the vacuum Lamb shift
which may be omitted, since the (renormalized) vacuum Lamb 
shift may be thought of as being included in the atomic 
transition frequency $\omega_{\rm A}$. In this way, 
Eq.~(\ref{e19.2}) takes the form 
\begin{eqnarray}
\label{e19.5}
\lefteqn{
        \bar{K}(t-t') = - {\textstyle\frac{1}{2}}A_0 + 
        \frac{k_{\rm A}^2 \mu_i \mu_j}{\hbar\pi\epsilon_0}
}
\nonumber\\&&\hspace{2ex}\times         
         \int_0^\infty \!\!d\omega\,
        \frac{{\rm Im}\,G^{\rm R}_{ij}({\bf r}_{\rm A},{\bf r}_{\rm A},\omega)}
        {i(\omega-\omega_{\rm A})}
        \left[e^{-i(\omega-\omega_{\rm A})(t-t')}-1 \right].
\end{eqnarray}
The integral equation (\ref{e19.1}) together with the kernel
function (\ref{e19.5}) can be regarded as the basic equation
for studying the influence of an arbitrary configuration of
dispersing and absorbing dielectric matter on the spontaneous
decay of an excited atom. 


\subsection{Emission pattern}
\label{sec2.3}

The intensity of the light registered by a point-like
photodetector at position ${\bf r}$ and time $t$ is given by
\begin{equation}
\label{e24}
           I({\bf r},t) \equiv 
           \langle \psi(t) | 
           \hat{\bf E}^{(-)}({\bf r})\cdot\hat{\bf E}^{(+)}({\bf r})
           | \psi(t) \rangle \ .
\end{equation}
To obtain the emission pattern associated with the spontaneous decay 
of an excited atom in the presence of dispersing and absorbing 
dielectric matter, we combine Eqs.~(\ref{e10}), (\ref{e11}), (\ref{e8}),  
and (\ref{e15}). After some algebra we derive, on using Eq.~(\ref{e17a}),
\begin{eqnarray}
\label{e25}
\lefteqn{
          I({\bf r},t)  =
          \sum_i \biggl| {k_{\rm A}^2\mu_j \over \pi\epsilon_0}
          \int_0^t dt' \Big[ C_{u}(t')
}          
\nonumber \\ && \hspace{4ex} \times 
          \int_0^\infty d\omega\, 
          {\rm Im}\, G_{ij} ({\bf r},{\bf r}_{\rm A},\omega)
           e^{-i(\omega-\omega_{\rm A})(t-t')}
          \Big]\biggr|^2 ,
\end{eqnarray}
where we have again set $\omega$ $\!=$ $\!\omega_{\rm A}$
in the frequency integral. Again, all the relevant dielectric-matter
parameters are contained in the Green tensor.
In contrast to Eq.~(\ref{e19.1}) together with the kernel
function (\ref{e19.5}), Eq.~(\ref{e25}) requires information 
about the Green tensor at different space points.
In particular, its dependence on space and frequency 
essentially determines the retardation effects.

In the simplest case of the atom being in free space we have
\begin{eqnarray}
\label{e25a}
\lefteqn{
         \mu_j{\rm Im}\,G^{\rm V}_{ij} ({\bf r},{\bf r}_{\rm A},\omega)
         = {1\over 8i\pi\rho} \left( \bbox{\mu}
         - \frac{\bbox{\rho}\,\bbox{\rho}\cdot\bbox{\mu}}
         {\rho^2} \right)_i
}         
\nonumber\\&&\hspace{10ex}\times\,
         \left(e^{i\omega\rho/c} - e^{-i\omega\rho/c} \right) 
         + {\cal O}\!\left(\rho^{-2}\right)
\end{eqnarray}
($\bbox{\rho}$ $\!=$ $\!{\bf r}$ $\!-$ $\!{\bf r}_{\rm A}$).
We substitute the expressions (\ref{e20}) (with $A$ $\!=$ $\!A_0$)
and (\ref{e25a}) into Eq.~(\ref{e25}), calculate the time integral, 
and extend in the frequency integral the lower limit to $-\infty$, 
which then can be calculated by contour integration,
\begin{eqnarray}
\label{e25b}
\lefteqn{
         \int_{-\infty}^\infty d\omega 
         \left(e^{i\omega\rho/c} - e^{-i\omega\rho/c} \right)
         \frac
         {e^{-(A_0/2+i\omega_{\rm A}')t}
         - e^{-i\omega t} }
         {i[\omega - (\omega_{\rm A}'- iA_0/2)]}
}         
\nonumber\\[.5ex]
         &&\hspace{3ex}
         =\, - 2\pi \exp\!\left[(-{\textstyle\frac{1}{2}}A_0  
         - i\omega_{\rm A}') (t-\rho/c)\right]
         \Theta(t-\rho/c) ,
\end{eqnarray}
where
\begin{eqnarray}         
\label{e25c}
         \omega_{\rm A}' = \omega_{\rm A}
         - {\textstyle\frac{1}{2}} \delta\omega 
\end{eqnarray}
[$\Theta(x)$, unit step function]. We thus derive the
well-known (far-field) result that
\begin{equation}
\label{e25d}
          I({\bf r},t)  =
          \left( {k_{\rm A}^2\mu\sin\theta \over 
          4\pi\epsilon_0 \rho} \right)^2 
          e^{-A_0(t-\rho/c)} \, \Theta(t-\rho/c),
\end{equation}
where $\theta$ is the angle between $\bbox{\rho}$ and
$\bbox{\mu}$.

Let us return to the general expression (\ref{e25}). If retardation 
is ignored and the Markov approximation applies, then we can replace, 
for all ${\bf r}$, $C_{u}(t')$ by $C_{u}(t)$ in the time integral in 
Eq.~(\ref{e25}) and approximate the time integral by 
$\zeta(\omega_A$ $\!-$ $\!\omega)$. We obtain
\begin{equation}
\label{e26}
          I({\bf r},t)  = 
          |{\bf F}({\bf r},{\bf r}_{\rm A},\omega_{\rm A})|^2 
          e^{-At} ,
\end{equation}
where
\begin{eqnarray}
\label{e26a}
\lefteqn{
          F_i({\bf r},{\bf r}_{\rm A},\omega_{\rm A}) = 
          {k_{\rm A}^2\mu_j \over \pi\epsilon_0}
}
\nonumber\\&&\hspace{6ex}\times\,          
          \!\int_0^\infty \!\!d\omega  \,
          {\rm Im}\, G_{ij} ({\bf r},{\bf r}_{\rm A},\omega)
          \zeta(\omega_{\rm A}-\omega).
\end{eqnarray}


\subsection{Emitted-light spectrum}
\label{sec2.4}

Next, let us consider the time-dependent power spectrum of 
the emitted light, which for sufficiently small passband width
of the spectral apparatus can be given by (see, e.g., \cite{22a})
\begin{eqnarray}
\label{e26.1}
\lefteqn{
          S({\bf r},\omega_{\rm S},T) = 
          \int_0^T dt_2\int_0^T dt_1 \Big[e^{-i\omega_{\rm S}(t_2-t_1)}
}
\nonumber\\&&\hspace{19ex}\times\,          
          \langle \hat{\bf E}^{(-)}({\bf r},t_2)\cdot 
          \hat{\bf E}^{(+)}({\bf r},t_1) \rangle\Big], 
\end{eqnarray}
where $\omega_{\rm S}$ is the setting frequency of the spectral 
apparatus and $T$ is the operating-time interval of the detector.
In close analogy to the derivation of Eq.~(\ref{e25}),
we combine Eqs.~(\ref{e10}), (\ref{e11}), (\ref{e8}),  
and (\ref{e15}) and use the relation (\ref{e17a}) to obtain
\begin{eqnarray}
\label{e26.3}
\lefteqn{
          S({\bf r},\omega_{\rm S},T) = 
          \sum_i \biggl| {k_{\rm A}^2\mu_j \over \pi\epsilon_0}
          \!\!\int_0^T \!\!dt_1\Big[ 
          e^{i(\omega_{\rm S}-\omega_{\rm A}) t_1}
          \!\!\int_0^{t_1}\!\! dt'\,C_{u}(t')
}
\nonumber\\&&\hspace{6ex}\times
          \int_0^\infty\!\! d\omega\, 
          {\rm Im}\, G_{ij} ({\bf r},{\bf r}_{\rm A},\omega)
          e^{-i(\omega-\omega_{\rm A}) (t_1-t')}
          \Big]\biggr|^2 .          
\end{eqnarray}
Further calculation again requires
knowledge of the Green tensor of the problem.

Let us apply Eq.~(\ref{e26.3}) to the free-space case. Following
the line that has led from Eq.~(\ref{e25}) to Eq.~(\ref{e25d}), 
we find that 
\begin{eqnarray}
\label{e26.3a}
\lefteqn{
           S({\bf r},\omega_{\rm S},T) = 
          \left( {k_{\rm A}^2 \mu \sin\theta \over
          4\pi \epsilon_0\rho} \right)^2
}          
\nonumber\\&&\hspace{2ex}\times\,
          \left| \frac
          {e^{[-A_0/2+i(\omega_{\rm S}-\omega_{\rm A}')](T-\rho/c)} -1}
          {\omega_{\rm S}- \omega_{\rm A}'+iA_0/2} \right|^2
          \Theta(T-\rho/c) .
\end{eqnarray}
In particular for $T$ $\!\rightarrow$ $\!\infty$, we recognize
the well-known Lorentzian:
\begin{eqnarray}
\label{e26.3b}
\lefteqn{
          \lim_{T\to\infty} S({\bf r},\omega_{\rm S},T) = 
          \left( {k_{\rm A}^2 \mu \sin\theta \over
          4\pi \epsilon_0\rho} \right)^2
}         
\nonumber\\[.5ex]&&\hspace{10ex}\times\,
          {1\over 
          [\omega_{\rm S} - (\omega_{\rm A}-\delta\omega/2)]^2
          +A_0^2/4} \, .
\end{eqnarray}

When retardation is ignored and the Markov approximation applies,
then Eq.~(\ref{e26.3}) can be simplified in a similar
way as Eq.~(\ref{e25}). In close analogy to the derivation of 
Eq.~(\ref{e26}) we may write
\begin{eqnarray}
\label{e26.4}
\lefteqn{
          S({\bf r},\omega_{\rm S},T) = 
          |{\bf F}({\bf r},{\bf r}_{\rm A},\omega_{\rm A})|^2
}          
\nonumber\\ &&\hspace{6ex}\times\, 
          \left| \frac
          {e^{\{-A/2+i[\omega_{\rm S}
          -(\omega_{\rm A}-\delta\omega/2)]\} T} -1} 
          {\omega_{\rm S}-(\omega_{\rm A}-\delta\omega/2)
          + iA/2} \right|^2 ,
\end{eqnarray}
with ${\bf F}({\bf r},{\bf r}_{\rm A},\omega_{\rm A})$ being defined
by Eq.~(\ref{e26a}).


\section{Application to a spherical cavity}
\label{sec3}

\subsection{The model}
\label{sec3.1}

\begin{figure}[!t!]
\noindent
\begin{center}
\epsfig{figure=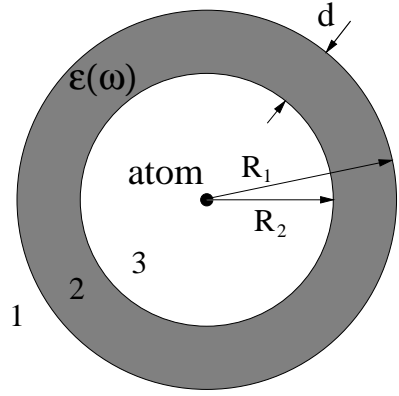,width=0.6\linewidth}
\end{center}
\caption{ 
Scheme of the spherical cavity.
}
\label{fig1}
\end{figure}

We apply the formalism developed in Sec.~\ref{sec2} to the 
spontaneous decay of an excited two-level atom placed inside  
a spherical three-layered structure (Fig.~\ref{fig1}). 
The outer layer
($r$ $\!>$ $\!R_1$) and the inner layer ($0$ $\!\le$ $\!r$ $\!<$ $\!R_2$)
are vacuum, while the middle layer ($R_2$ $\!\le$ $\!r$ $\!\le$ $\!R_1$)
is a dispersing and absorbing dielectric. The Green tensor of
the configuration is given in Appendix \ref{appB}.

\begin{figure}[!t!]
\noindent
\begin{center}
\epsfig{figure=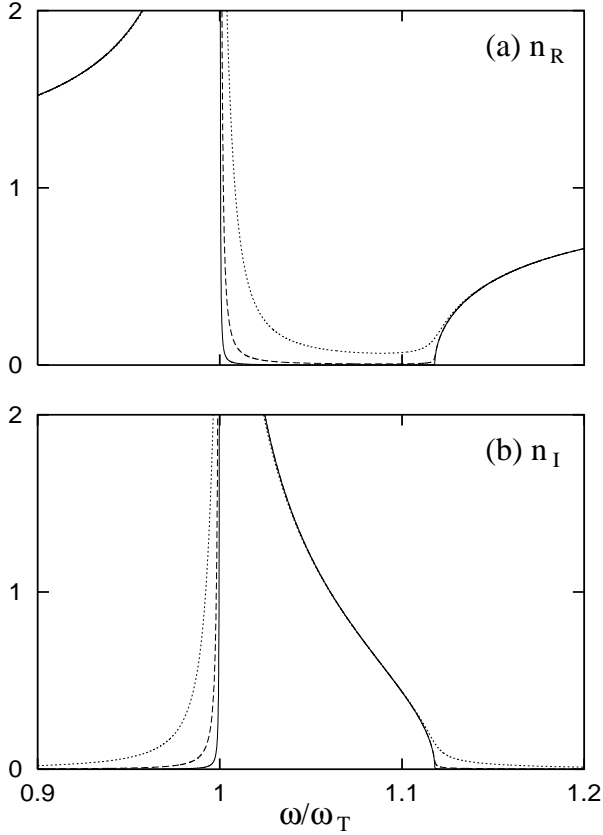,width=1.\linewidth}
\end{center}
\caption{
(a) The real part $n_{\rm R}(\omega)$ and 
(b) the imaginary part $n_{\rm I}(\omega)$ of the 
complex refractive index are shown for 
$\omega_{\rm P}$ $\!=$ $\!0.5\,\omega_{\rm T}$ and  
$\gamma$ $\!=$ $\!10^{-4}\,\omega_{\rm T}$ (solid line), 
$\gamma$ $\!=$ $\!10^{-3}\,\omega_{\rm T}$ (dashed line), and 
$\gamma$ $\!=$ $\!10^{-2}\,\omega_{\rm T}$ (dotted line). 
The longitudinal frequency $\omega_{\rm L}$ $\!=$ 
$\!\protect\sqrt{\omega_{\rm T}^2\!+\!\omega_{\rm P}^2}$ 
is $\omega_{\rm L}$ $\!\simeq$ $\!1.12\,\omega_{\rm T}$, and
hence $\omega_{\rm L}$ $\!-$ $\!\omega_{\rm T}$
$\!\simeq$ $\!0.12\,\omega_{\rm T}$. 
}
\label{fig2}
\end{figure}

We have performed the calculations assuming a 
Lorentz-type dielectric with a single resonance  
(in the relevant frequency region):
\begin{equation}
\label{e30}
        \epsilon(\omega) = 1 + 
        {\omega_{\rm P}^2 \over 
        \omega_{\rm T}^2 - \omega^2 - i\omega \gamma}\,.
\end{equation}
Here, $\omega_{\rm P}$ is the plasma frequency, which
is proportional to the square root of the number density of the Lorentz 
oscillators and plays the role of the coupling constant between the 
medium polarization and the electromagnetic field, and $\omega_{\rm T}$ 
and $\gamma$, respectively, are the position and the width of the 
medium resonance. 
The plots in Fig.~\ref{fig2} of the real and imaginary parts of the index 
of refraction
\begin{equation}
\label{e30a}
n(\omega) = \sqrt{\epsilon(\omega)}
          = n_{\rm R}(\omega) + i n_{\rm I}(\omega) 
\end{equation} 
illustrate a typical band-gap behavior of the configuration.

\subsection{Weak-coupling regime}
\label{sec3.2}

In the weak-coupling regime, the excited atomic state decays
exponentially [Eq.~(\ref{e20})]. For simplicity let us assume that 
the atom is positioned at the center of the cavity. From Eqs.~(\ref{e21}), 
(\ref{e19.3}), (\ref{e19.4}), and (\ref{B24}), the cavity-modified 
decay rate is then found to be
\begin{equation}
\label{e34a}
         A = \bar A(\omega_A)\,A_0\,,
\end{equation}
where $A_0$ is decay rate in free space, Eq.~(\ref{e19.6}), and
\begin{equation}
\label{e34}
         \bar{A}(\omega) = 
         1+{\rm Re}\, C^{33}_N (\omega) \ ,
\end{equation}
with $C^{33}_N (\omega)$ being given by 
Eqs.~(\ref{B6}) -- (\ref{B21}). 
Note that if mode expansion applies, $\bar A(\omega)$ would 
correspond to the change of the density of modes
due to the presence of the cavity.

When far from the medium resonance absorption is disregarded 
and hence the (frequency-independent) refractive index is assumed 
to be real, then previous results obtained by mode decomposition 
can be recovered. In particular, for $R_2$ $\!\to$ $\!0$ we 
recognize the decay rate obtained in \cite{29,33} for 
microspheres and liquid droplets. However, it should be pointed 
out that even far from the medium resonance the imaginary
part of the refractive index cannot be set equal to zero in general, 
since the contribution to the decay rate of the nonradiative 
decay associated with absorption increases $\sim$ $\!R_2^{-3}$ 
for decreasing $R_2$ (and nonvanishing imaginary part
of the refractive index) \cite{9}. 

\begin{figure}[!t!]
\noindent
\begin{center}
\epsfig{figure=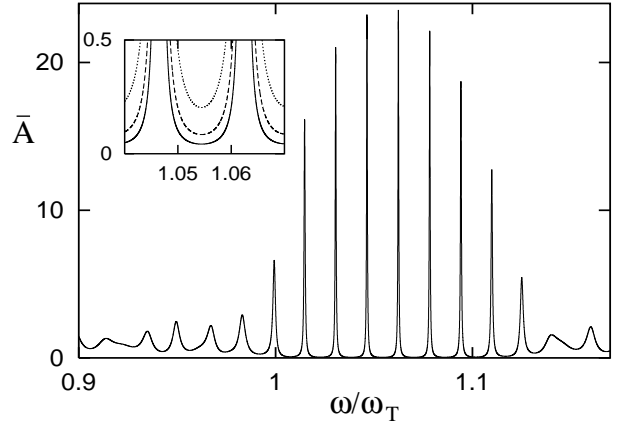,width=1\linewidth}
\end{center}
\caption{ 
The function $\bar A(\omega)$, Eq.~(\protect\ref{e34}), is shown for
$R_2$ $\!=$ $\!30\,\lambda_{\rm T}$,
$d$ $\!=$ $\!\lambda_{\rm T}$,
$\omega_{\rm P}$ $\!=$ $\!0.5\,\omega_{\rm T}$, 
and $\gamma$ $\!=$ $\!10^{-2}\,\omega_{\rm T}$.
The curves in the inset correspond to
$\gamma=10^{-2}\,\omega_{\rm T}$ (solid line),  
$\gamma=2\times 10^{-2}\,\omega_{\rm T}$ (dashed line), and
$\gamma=5\times 10^{-2}\,\omega_{\rm T}$ (dotted line).
}
\label{fig3}
\end{figure}
 
Let us restrict our attention to a true microcavity 
($R_2\omega_{\rm A}/c$ $\!\gg$ $\!1$). From Fig.~\ref{fig3} it is seen 
that the rate of spontaneous decay sensitively depends 
on the transition frequency. Narrow-band enhancement 
of spontaneous decay \mbox{($\bar A$ $\!>$ $\!1$)} 
alternates with broadband inhibition ($\bar A$ $\!<$ $\!1$). 
The frequencies where the maxima of enhancement are observed 
correspond to the resonance frequencies of the cavity.
Within the band gap the heights and widths of the 
frequency intervals in which spontaneous decay is feasible 
are essentially determined by the material losses. 
Outside the band-gap zone the change of the decay rate is less 
pronounced because of the relatively large input--output coupling,
the (small) material losses being of secondary importance.   

When $R_2\omega/c$ $\!\gg$ $\!1$ and 
$\exp[-n_{\rm I}(R_1$ $\!-$ $\!R_2)\omega/c]$ $\!\ll$ $\!1$, then
Eqs.~(\ref{B6}) -- (\ref{B21}) drastically simplify and $\bar A(\omega)$ 
[Eq.~(\ref{e34})] reads 
\begin{equation}
\label{e36}
        \bar{A}(\omega) \simeq {\rm Re}\! \left[ \frac
        {n(\omega)- i\tan (R_2 \omega/c)} 
        {1 - i n(\omega) \tan (R_2 \omega/c)} \right] .
\end{equation}
The positions $\omega_m$ of the maxima of $\bar{A}(\omega)$ can then
be obtained from the equation $d\bar A/d\omega$ $\!=$
$\!0$. As long as $n(\omega)$ can be regarded as being
slowly varying compared with $\tan (\omega R_2/c)$, 
we may neglect $dn/d\omega$ and thus determine
the resonance frequencies from the equation
\enlargethispage{4mm}
\begin{samepage}
\begin{eqnarray}
\label{e38}
\lefteqn{
        2 n_{\rm I}(\omega_m)\, \tan(R_2 \omega_m /c )   
}
\nonumber 
\\&&\hspace{2ex}        
        \simeq\, |n(\omega_m)|^2 \!-\! 1\! 
        - \!\sqrt{(|n(\omega_m)|^2\!-\!1)^2\!+\!4n_{\rm I}^2(\omega_m)}\,. 
\end{eqnarray}
\end{samepage}
Note that Eq.~(\ref{e38}) is exact when it is regarded
as conditional equation of $R_2$ for a desired resonance frequency.

In the band-gap zone we may assume that $n_{\rm I}$ $\!\gg$ $\!n_{\rm R}$
(see Fig.~\ref{fig2}). From Eqs.~(\ref{e36}) and (\ref{e38}) it then
follows that the maximum values $\bar{A}(\omega_m)$ and
half widths at half maximum, $\delta \omega_m$, of the
cavity resonance lines, i.e., the regions where enhanced
spontaneous decay can be observed, are given by 
\begin{equation}
\label{e39}
         \bar{A}(\omega_m) \simeq
         { n_{\rm I}^2(\omega_m)+1\over n_{\rm R}(\omega_m)} \simeq 
         \frac{2\sqrt{(\omega_{\rm L}^2-\omega_m^2)
         (\omega_m^2-\omega_{\rm T}^2)}}
         {\gamma\omega_m} \,,
\end{equation}
\begin{equation}         
\label{e39a}
         \delta \omega_m \simeq {c \over R_2 \bar{A}(\omega_m)} \,, 
\end{equation}
where we have assumed that the material losses are small, i.e.,
$\gamma$ $\!\ll$ $\!\omega_{\rm T},\,\omega_{\rm P},\,
\omega_{\rm P}^2/\omega_{\rm T}$. Equations (\ref{e39}) and 
(\ref{e39a}) reveal that in the approximation made the 
heights (widths) of the resonance lines (Lorentzians) are 
proportional (inversely proportional) to $\gamma$, the highest 
and narrowest line being in the center of the band gap.
Note that the product $\bar{A}(\omega_m)\delta \omega_m$ does not 
depend on $\gamma$.

Outside the band-gap zone the inequality $n_{\rm R}$ $\!\gg$ $\!n_{\rm I}$ 
is typically valid (see Fig.~\ref{fig2}), and thus Eqs.~(\ref{e36}) and 
(\ref{e38}) yield 
\begin{equation}
\label{e41}
         \bar{A}(\omega_m) \simeq n_{\rm R}(\omega_m) 
         \simeq \sqrt{\frac {\omega_{\rm L}^2-\omega_m^2}
            {\omega_{\rm T}^2-\omega_m^2}}\,,
\end{equation} 
\begin{equation}
\label{e41a}           
         \delta \omega_m \simeq {c \over R_2 \bar{A}(\omega_m)}\,.
\end{equation}
The heights and widths of the resonance lines are now seen to be
(approximately) independent of $\gamma$. The lines become higher 
and narrower if $\omega_m$ becomes close to $\omega_{\rm T}$.

\begin{figure}[!t!]
\noindent
\begin{center}
\epsfig{figure=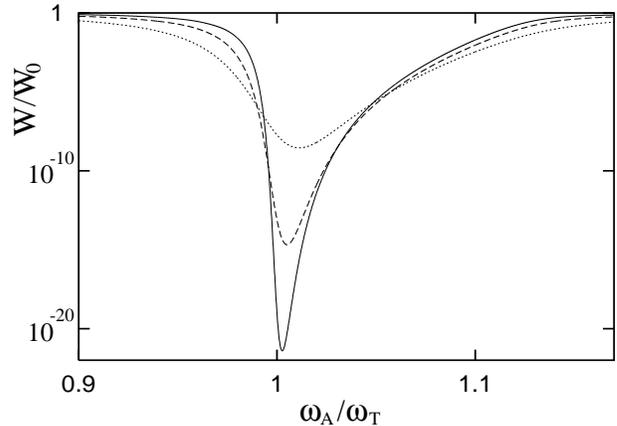,width=1\linewidth}
\end{center}
\caption{ 
The amount of radiation energy $W$, Eq.~(\protect\ref{E41C}),
observed outside the cavity
is shown as a function of the atomic transition frequency
$\omega_{\rm A}$ for
$\gamma=10^{-2}\,\omega_{\rm T}$ (solid line),  
$\gamma=2\times 10^{-2}\,\omega_{\rm T}$ (dashed line), and
$\gamma=5\times 10^{-2}\,\omega_{\rm T}$ (dotted line).
The other parameters are the same as in Fig.~\protect\ref{fig3}.
}
\label{fig3-1}
\end{figure}

The widths of the resonance lines are responsible for the damping
of intracavity fields. There are two damping mechanisms:
photon leakage to the outside of the cavity and 
photon absorption by the cavity-wall material. From the analysis 
given above it is seen that 
the first mechanism is the dominant one outside the band gap where
normal dispersion \mbox{($dn_{\rm R}/d\omega$ $\!>$ $\!0$)} 
is observed, while the latter dominates inside the band gap where 
anomalous dispersion \mbox{($dn_{\rm R}/d\omega$ $\!<$ $\!0$)} 
is observed. To illustrate this in more detail, we have calculated
the amount of radiation energy observed outside the cavity,
\begin{equation}
\label{e41b}
W = 2c\epsilon_0\int_0^\infty dt \int_0^{2\pi} d\phi
       \int_0^\pi d\theta\,  \rho^2 \sin\theta \,I({\bf r},t) 
\end{equation}
($\rho$ $\!>$ $\!R_1$), and compared it with the emitted energy 
in free space $W_0$ $\!=$ $\!\hbar\omega_{\rm A}$. Assuming without 
loss of generality that the atomic transition dipole is $z$-oriented
and restricting our attention to the relevant 
far-field contribution, from Eqs.~(\ref{e26}) and (\ref{e26a}) 
together with Eq.~(\ref{B1}) and Eqs.~(\ref{B25}) -- (\ref{B27}) 
we derive (see Appendix \ref{appC})
\begin{equation}
\label{E41C}
        {W\over W_0}  
        \simeq \frac{ |A^{13}_N(\omega_{\rm A})|^2}
        {1+{\rm Re} C^{33}_N(\omega_{\rm A}) }\,.
\end{equation}
Examples of the dependence on the atomic transition 
frequency of the amount of radiation energy observed outside 
the cavity are plotted in Fig.~\ref{fig3-1}. It is seen
that inside the gap most of the energy emitted by the
atom is absorbed by the cavity wall in the course of time, 
while outside the gap 
the absorption is (for the chosen values of $\gamma$) much 
less significant. Note that with increasing value of $\gamma$
the band gap is smoothed a little bit, and thus the
fraction of light that escapes from the cavity can increase.  

\subsection{Strong-coupling regime}
\label{sec3.3}

The strength of the atom-field coupling increases when the 
atomic transition frequency $\omega_{\rm A}$ approaches a cavity-resonance 
frequency $\omega_m$. In order to gain insight into the strong-coupling 
regime, let us first consider the limiting case of one
cavity-resonance line being involved in the atom-field
interaction. Indeed, when $\omega_{\rm A}$ is close to 
$\omega_m$, then contributions from the other 
resonance lines become small. Using Eqs.~(\ref{e21}), (\ref{e34a}),
and (\ref{e34}), and recalling that $\bar A(\omega)$ behaves
like a Lorentzian in the vicinity of $\omega_m$, we may
simplify Eq.~(\ref{e19}) to
\begin{eqnarray}
\label{e41.1}
\lefteqn{
        K(t-t') 
        \simeq - {A_0 \over 2\pi} \, \bar{A}(\omega_m) 
        (\delta\omega_m)^2
        e^{-i(\omega_m-\omega_{\rm A})(t-t')}        
}        
\nonumber\\ [.5ex]
        &&\hspace{15ex}\times\,
        \int_{-\infty}^{+\infty} \!\!d\omega\,
        \frac{e^{-i(\omega-\omega_m)(t-t')}}
        {(\omega-\omega_m)^2 + (\delta\omega_m)^2}
\nonumber\\ [.5ex] 
        &&\hspace{3ex}
        = -{\textstyle\frac{1}{2}} A_0 \bar A(\omega_m) \delta\omega_m
        e^{-i(\omega_m-\omega_{\rm A})(t-t')} e^{- \delta\omega_m |t-t'|}\,.
\end{eqnarray}
Substituting this expression into Eq.~(\ref{e18})
and differentiating both sides of the resulting equation
with regard to time, we arrive at
\begin{equation}
\label{e41.2}
         \ddot{C}_u(t) 
         + \left[ i(\omega_m\!-\!\omega_{\rm A}) + \delta\omega_m \right]
         \dot{C}_u(t) + (\Omega/2)^2 C_{u}(t) = 0
\end{equation}
where
\begin{equation}
\label{e41.3}
         \Omega = \sqrt{2A_0\bar{A}(\omega_m) \delta\omega_m} \,.
\end{equation}
Hence we are left, in the approximation made, with a 
damped-oscillator equation of motion for the upper-state
probability amplitude, where $\bar{A}(\omega_m)$ and $\delta\omega_m$, 
respectively, are given by Eqs.~(\ref{e39}) and (\ref{e39a}) [or
Eqs.~(\ref{e41}) and (\ref{e41a})]. 
Obviously, when $\omega_{A}$ $\!=$ $\omega_m$ and 
\begin{equation}
\label{e41.4}
         \Omega \gg \delta\omega_m 
\end{equation}
(i.e., strong coupling), then damped Rabi oscillations are observed: 
\begin{equation}
\label{e41.5}
\left| C_{u}(t) \right|^2 = e^{-\delta\omega_m t} 
\cos^2(\Omega t/2) .
\end{equation}
Note that in the opposite case where $\Omega$ $\!\ll$ $\!\delta\omega_m$
the solution of Eq.~(\ref{e41.2}) is $|C_{u}(t)|^2 $
$\!=$ $\!e^{-A t}$ with $A$ from Eq.~(\ref{e34a}) for 
$\omega_A$ $\!=$ $\!\omega_m$, which is in agreement with Eq.~(\ref{e20}).
Equations (\ref{e41.3}) and (\ref{e41.4}) together with Eqs.~(\ref{e39}) and
(\ref{e39a}) [or Eqs.~(\ref{e41}) and (\ref{e41a})]
provide us with an easy rule of thumb for deciding 
whether the strong-coupling regime is realized or not.

\begin{figure}[!t!]
\noindent
\begin{center}
\epsfig{figure=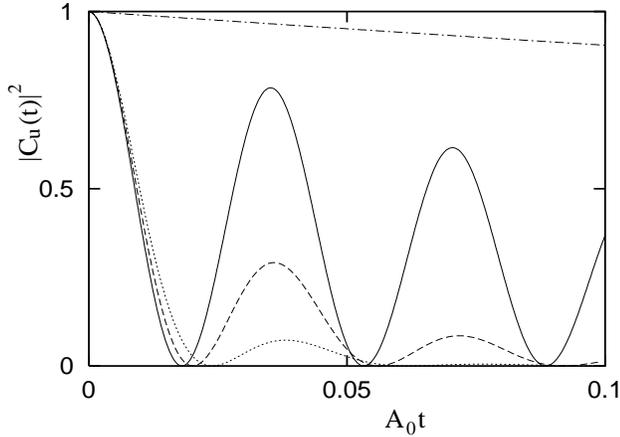,width=1\linewidth}
\end{center}
\caption{
The temporal evolution of the occupation probability
$|C_{u}(t)|^2$ of the upper atomic state is shown for
$R_2$ $\!=$ $\!30\,\lambda_{\rm T}$,
$d$ $\!=$ $\!\lambda_{\rm T}$,
$\omega_{\rm P}$ $\!=$ $\!0.5\,\omega_{\rm T}$, 
$\omega_{\rm A}$ $\!=$ $\!1.046448\omega_{\rm T}$, 
$A_0\lambda_{\rm T}/(2c)$ $\!=$ $\!10^{-6}$, and 
$\gamma$ $\!=$ $\!10^{-4}\,\omega_{\rm T}$ (solid line),  
$\gamma$ $\!=$ $\!5\times10^{-4}\,\omega_{\rm T}$ (dashed line), 
$\gamma$ $\!=$ $\!10^{-3}\,\omega_{\rm T}$ (dotted line).
For comparison, the exponential decay in free space is shown
(dashed-dotted line). 
}
\label{fig4}
\end{figure}
\begin{figure}[!t!]
\noindent
\begin{center}
\epsfig{figure=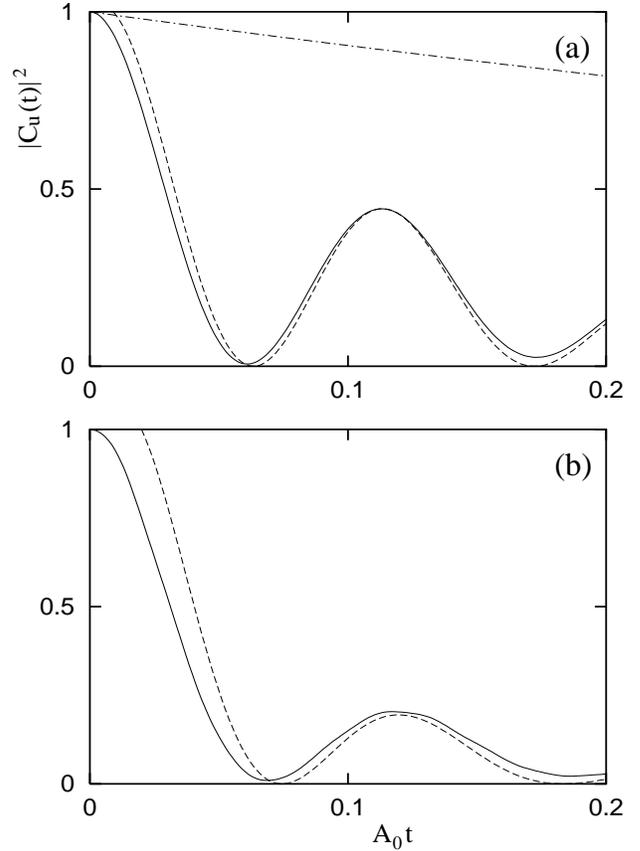,width=1\linewidth}
\end{center}
\caption{
The temporal evolution of the occupation probability
$|C_{u}(t)|^2$ of the upper atomic state is shown for
$d$ $\!=$ $\!\lambda_{\rm T}$,
$\omega_{\rm A}$ $\!=$ $\!0.9999\,\omega_{\rm T}$,
$A_0\lambda_{\rm T}/(2c)$ $\!=$ $\!10^{-5}$, and 
(a) $\omega_{\rm P}$ $\!=$ $\!3\,\omega_{\rm T}$, 
$R_2$ $\!=$ $\!30.00197\,\lambda_{\rm T}$,
(b) $\omega_{\rm P}$ $\!=$ $\!1.5\,\omega_{\rm T}$,
$R_2$ $\!=$ $\!30.00179\,\lambda_{\rm T}$.
The solid lines correspond to the exact solution
and the dashed lines to the approximate analytical 
solution [solution of Eq.~(\protect\ref{e41.2})] 
shifted forwards by 
(a) $A_0\Delta$ $\!t=$ $\!0.009$ and
(b) $A_0\Delta$ $\!t=$ $\!0.02$.
For comparison, the exponential decay in free space is 
shown (dashed-dotted line).
}
\label{fig5}
\end{figure}

In order to obtain the exact solution to the problem, 
we have also solved the basic integral equation (\ref{e19.1}) 
[together with the kernel function (\ref{e19.5})] numerically.
Typical examples of the temporal evolution of the occupation
probability of the upper atomic state are shown in
Fig.~\ref{fig4} for the case where the atomic transition 
is tuned to the resonance line in the center of the band gap. 
The figure reveals that with increasing value
of the intrinsic damping constant $\gamma$ of the wall material
the Rabi oscillations become less pronounced, in agreement
with Eqs.~(\ref{e39}), (\ref{e39a}), (\ref{e41.3}), and
the inequality (\ref{e41.4}). From Eqs.~(\ref{e39}) and (\ref{e39a}) 
it is seen that (in the approximation made there) the product 
$\bar A(\omega_m) \delta\omega_m$ $\!\simeq$ $\!c/R_2$ does not 
vary with $\gamma$, whereas $\delta\omega_m$ linearly increases 
with $\gamma$. According to Eq.~(\ref{e41.3}), \mbox{$\Omega$ $\!\propto$ 
$\!\sqrt{\gamma}$} and thus $\Omega/\delta\omega_m$ $\!\propto$
$\!1/\sqrt{\gamma}$,  i.e., the condition of strong coupling (\ref{e41.4})
becomes violated  with increasing $\gamma$. 
Physically, increasing $\gamma$ means increasing probability 
of (irreversible) absorption of the emitted photon by the wall material 
and therefore reduced probability of (reversible) atom-field energy exchange. 
It should be mentioned that with increasing $\gamma$ the (very small) 
probability that the photon (irreversibly) leaves the cavity also increases.
Note that the variation of $\gamma$ in Fig.~\ref{fig4} leaves the imaginary
part of the refractive index nearly unchanged, $n_{\rm I}$ $\!\simeq$
$\!1.2$, while the (small) real part $n_{\rm R}$ slightly 
increases with $\gamma$ (see Fig.~\ref{fig2}).

The examples of the temporal evolution of the occupation
probability of the upper atomic state shown in
Fig.~\ref{fig5} refer to the case where the atomic transition 
is tuned to a resonance line closest to $\omega_{\rm T}$
below the band gap. According to Eqs.~(\ref{e41}), (\ref{e41a}), 
(\ref{e41.3}), and the condition (\ref{e41.4}), the resonance 
frequencies close to $\omega_{\rm T}$ are most favorable for 
realizing the strong-coupling regime in the range of normal dispersion, 
because of the rising real part of the refractive index
[see Fig.~\ref{fig2}(a)]. As expected, the strength of the the Rabi 
oscillations now varies with the plasma frequency $\omega_{\rm P}$ such 
that they are less pronounced for small values of $\omega_{\rm P}$. 
Obviously, decreasing $\omega_{\rm P}$ means increasing input-output 
coupling, i.e., increasing probability that the emitted photon leaves 
the cavity instead of back-acting upon the atom. 

Finally, Fig.~\ref{fig5} presents a comparison between the exact solution,
and the approximate analytical solution [solution of
Eq.~(\ref{e41.2})]. To compensate for the
short-time inaccuracy of the analytical solution, it is somewhat
shifted forward in time. The agreement between the
exact solution and the (shifted) analytical solution 
is quite good. Obviously, Eq.~(\ref{e41.2}) predicts an 
initial decay somewhat faster than the exact one. 
In fact, the spontaneous decay is accelerated
only  gradually under the back-action of the radiated
field being multiply reflected at the boundaries \cite{31,18},
and the single-resonance coupling 
is established only after a certain interval of time. 
The sharper is the cavity resonance, the shorter is this interval, 
which is fully confirmed by the figure. 


\section{Summary and Conclusions}
\label{sec4}

We have developed a formalism for studying spontaneous decay
of an excited atom in the presence of arbitrary dispersing and
absorbing dielectric bodies. The formalism is based on a 
source-quantity representation of the electromagnetic field
in terms of the Green tensor of the classical problem and
appropriately chosen bosonic quantum fields. It
replaces the standard concept of mode decomposition which fails 
for complex permittivity. All relevant information about the 
bodies such as form and intrinsic dispersion and absorption
properties are contained in the Green tensor.
It is worth noting that the Green tensor has been available 
for a large variety of configurations such as planarly, spherically, 
and cylindrically multilayered media \cite{22}.  

We have applied the theory to the spontaneous decay
of a two-level atom placed at the center of a three-layered
spherical microcavity, modeling the wall by a Lorentz dielectric. 
The formalism has enabled us to study both the 
range of normal dispersion and the anomalous-dispersion
range within the band gap in a unified way. Whereas
in the range of normal dispersion the cavity input--output 
coupling dominates the strength of the atom-field interaction,
the dominating effect within the band gap is the
photon absorption by the wall material.

In the study of the spherical-cavity problem, we have
assumed that the atom is placed at the center of the
cavity, which has drastically reduced the mathematical effort,
because only the spherical Bessel function of order $n$ $\!=$ 
$\!1$ contributes to the Green tensor. 
The price we paid is that the interaction of the atom 
with cavity excitations
of large $n$, which correspond to high-$Q$ whispering 
gallery modes and concentrate near the surface by repeated 
internal reflections has not been included in the
analysis. Since the complete Green tensor is known,
there is of course no obstacle to perform the calculations
for an arbitrary position of the atom. In particular, when
the atom is near the surface then nonradiative energy
transfer from the atom to the absorbing medium can
substantially contribute to the process of spontaneous decay.  
Further investigations are also necessary in order to give
a more detailed analysis of the evolution of the emitted radiation,
to answer the question of the ratio of photon emission
to nonradiative decay, and to extend the theory to multilevel
atom-field interactions. 

\acknowledgements

We thank Stefan Scheel for helpful discussions. H.T.D. 
is grateful to the Alexander von Humboldt-Stiftung
for a research fellowship. This work was supported by the 
Deutsche Forschungsgemeinschaft.


\appendix

\section{The Hamiltonian in the dipole and rotating wave 
approximations}
\label{appA}

The Hamiltonian (\ref{e12.5}) can be rewritten as
\begin{eqnarray}
\label{A1}
          &\displaystyle
          \hat{H} = \hat{H}_{\rm F} 
          +  \hat{H}_{\rm A} + \hat{H}_{\rm AF} ,
\\ [.5ex]
\label{A2}
          &\displaystyle
          \hat{H}_{\rm F} =
          \int d^3{\bf r} \int_0^\infty d\omega \,\hbar\omega\,
          \hat{\bf f}^\dagger({\bf r},\omega)\cdot 
          \hat{\bf f}({\bf r},\omega) ,
\\ [.5ex]
\label{A3}
          &\displaystyle
          \hat{H}_{\rm A} =
          \sum_\alpha {\hat{\bf p}_\alpha^2 \over 2m_\alpha} 
          + {1\over 2} \int d^3{\bf r} 
          \hat{\rho}_{\rm A}({\bf r}) 
          \hat{\varphi}_{\rm A}({\bf r}) ,
\\ [.5ex]
\label{A4}
          &\displaystyle
          \hat{H}_{\rm AF} =
          - \sum_\alpha {q_\alpha\over m_\alpha} 
          \hat{\bf p}_\alpha \cdot \hat{\bf A}({\bf r}_\alpha) 
          + \int d^3{\bf r} 
          \hat{\rho}_{\rm A}({\bf r}) \hat{\varphi}({\bf r}) ,
\end{eqnarray}
where we have ignored the small $\hat{\bf A}^2$ term.
Note that in the Coulomb gauge $[\hat{\bf p}_\alpha,\hat{\bf A}]=0$.
For a neutral atom with the nucleus being positioned at 
${\bf r}_{\rm A}$, the atomic dipole operator reads
\begin{equation}
\label{A5}
         \hat{\bbox{\mu}}_{\rm A} \equiv 
         \sum_\alpha q_\alpha (\hat{\bf r}_\alpha-{\bf r}_{\rm A})
         = \sum_\alpha q_\alpha \hat{\bf r}_\alpha \ ,
\end{equation}
and the first term in the interaction part of the Hamiltonian
$\hat{H}_{\rm AF}$ takes the form
\begin{eqnarray}
\label{A6}
          - \sum_\alpha {q_\alpha\over m_\alpha} 
          \hat{\bf p}_\alpha \cdot \hat{\bf A}({\bf r}_\alpha) 
          &&\ \simeq\,
          - \sum_\alpha {q_\alpha\over i\hbar} 
          \left[\hat{\bf r}_\alpha,\hat{H}_{\rm A} \right] 
          \cdot \hat{\bf A}({\bf r}_{\rm A}) 
\nonumber\\
          &&\ =\,
          - {1\over i\hbar}
          \left[\hat{\bbox{\mu}}_{\rm A},\hat{H}_{\rm A} \right] 
          \cdot \hat{\bf A}({\bf r}_{\rm A}) ,
\end{eqnarray}
where the dipole approximation has been employed to replace 
$\hat{\bf A}({\bf r}_\alpha)$ $\!\rightarrow$ 
$\!\hat{\bf A}({\bf r}_{\rm A})$ and  
$\hat{\bf p}_\alpha$ $\!=$ $\!(m_\alpha/(i\hbar))
[\hat{\bf r}_\alpha,\hat{H}_{\rm A}]$ has been used.

Now we restrict ourselves to a two-state model of an atom  
with upper state $|u\rangle$ and lower state $|l\rangle$.
These are eigenstates of the unperturbed part of the 
Hamiltonian $\hat{H}_{\rm A}$ with the eigenvalues 
$\hbar \omega_u$ and $\hbar \omega_l$, respectively.
Then one can write
\begin{eqnarray}
\label{A7}
         &\displaystyle
         \hat{H}_{\rm A} = 
         \hbar \omega_u |u\rangle \langle u|
         + \hbar \omega_l |l\rangle \langle l|,
\\ [.5ex]
\label{A8}
         &\displaystyle
         |u\rangle \langle u|+|l\rangle \langle l|=\hat{I}.
\end{eqnarray}
In this atomic state space, the dipole 
operator $\hat{\bbox{\mu}}_{\rm A}$ has the matrix elements 
$\langle u| \hat{\bbox{\mu}}_{\rm A} |l\rangle =       
\langle l| \hat{\bbox{\mu}}_{\rm A} |u\rangle \equiv \bbox{\mu}$
and 
$\langle u| \hat{\bbox{\mu}}_{\rm A} |u\rangle =       
\langle l| \hat{\bbox{\mu}}_{\rm A} |l\rangle =0$. Using
them and Eqs.~(\ref{e12}), ({\ref{A6}) -- ({\ref{A8}), 
we arrive at
\begin{eqnarray}
\label{A9}
\lefteqn{
          - \sum_\alpha {q_\alpha\over m_\alpha}\, 
          \hat{\bf p}_\alpha \cdot \hat{\bf A}({\bf r}_\alpha) 
}
\nonumber\\ [.5ex] 
          &&\hspace{5ex}
          \simeq \left(\hat{\sigma} - \hat{\sigma}^\dagger \right)
          \left[ 
          \int_0^\infty d\omega\, {\omega_{\rm A}\over \omega}
          \underline{\hat{\bf E}}^\perp({\bf r}_{\rm A},\omega)
           - {\rm H.c.} \right] \cdot \bbox{\mu}
\nonumber\\[.5ex] 
          &&\hspace{5ex}\simeq 
          - \left[ \hat{\sigma}^\dagger 
          \hat{\bf E}^{\perp(+)}({\bf r}_{\rm A}) \cdot \bbox{\mu}
          + {\rm H.c.} \right] ,
\end{eqnarray}
where $\hat{\sigma}= |l\rangle \langle u|,\ 
\hat{\sigma}^\dagger = |u\rangle \langle l|$,  
$\omega_{\rm A}= \omega_u-\omega_l$, and $\omega=\omega_{\rm A}$
is set in the integral, because of the rotating wave approximation.

In order to deal with the second term in $\hat{H}_{\rm AF}$, we expand 
$\hat{\rho}_{\rm A}({\bf r})$ in a multi-polar form and retain
only the first non-vanishing term 
\begin{eqnarray}
\label{A10}
\lefteqn{
          \hat{\rho}_{\rm A}({\bf r})
          \simeq \sum_\alpha q_\alpha 
          \delta ({\bf r} -{\bf r}_{\rm A})
}
\nonumber\\[.5ex]
          &&\hspace{8ex}          
          - \bbox{\nabla} \cdot 
          \Big[ \delta ({\bf r} -{\bf r}_{\rm A})
          \sum_\alpha q_\alpha 
          (\hat{\bf r}_\alpha -{\bf r}_{\rm A}) \Big]
\nonumber\\[.5ex]
          && \hspace{6ex}=  
          - \bbox{\nabla}\cdot
          \delta ({\bf r} -{\bf r}_{\rm A})
          \hat{\bbox{\mu}}_{\rm A} \, .
\end{eqnarray}
Then we have
\begin{eqnarray}
\label{A11}
\lefteqn{
          \int d^3{\bf r}\, 
          \hat{\rho}_{\rm A}({\bf r}) \hat{\varphi}({\bf r})
          \simeq
          - \int d^3{\bf r}\, \big\{ \bbox{\nabla}\cdot
          [\delta({\bf r} -{\bf r}_{\rm A}) 
          \hat{\bbox{\mu}}_{\rm A}] \big\}
          \hat{\varphi}({\bf r})
}          
\nonumber\\[.5ex]&&\hspace{10ex}= 
          \int d^3{\bf r} [\delta({\bf r} -{\bf r}_{\rm A}) 
          \hat{\bbox{\mu}}_{\rm A}] \cdot \bbox{\nabla} 
          \hat{\varphi}({\bf r})
\qquad          
\nonumber\\[.5ex]&&\hspace{10ex}= 
          - \hat{\bbox{\mu}}_{\rm A}
          \cdot \hat{\bf E}^\parallel({\bf r}_{\rm A})
\qquad 
\nonumber\\[.5ex]&&\hspace{10ex}\simeq 
          - \left[ \hat{\sigma}^\dagger 
          \hat{\bf E}^{\parallel(+)}({\bf r}_{\rm A}) \cdot \bbox{\mu}
          + {\rm H.c.} \right] ,
\qquad 
\end{eqnarray}
where integration by parts and Eq.~(\ref{e12.1}) have been employed
for deriving the second and the third equation, respectively, 
and the rotating wave approximation
has been used in deriving the forth equation. 
Combining (\ref{A4}), (\ref{A9}),
and (\ref{A11}) gives
\begin{equation}
\label{A12}
          \hat{H}_{\rm AF} \simeq 
          - \left[ \hat{\sigma}^\dagger 
          \hat{\bf E}^{(+)}({\bf r}_{\rm A}) \cdot \bbox{\mu}
          + {\rm H.c.} \right] .
\end{equation}
Equations (\ref{A1}), (\ref{A2}), (\ref{A12}), and a subtraction
of $(\hbar/2)(\omega_u+\omega_l)\hat{I}$ from 
$\hat{H}_{\rm A}$, Eq.~(\ref{A7}), lead to the Hamiltonian (\ref{e14}).


\section{The Green tensor of the spherical cavity}
\label{appB}

Following\cite{22,35}, we write the Green tensor 
of the cavity in Fig.~\ref{fig1}
in the form
\begin{equation} 
\label{B1}
\bbox{G}({\bf r},{\bf r'},\omega) 
= \bbox{G}^{\rm V}({\bf r},{\bf r'},\omega) \delta_{fs} 
+ \bbox{G}^{(fs)}({\bf r},{\bf r'},\omega),
\end{equation}
where $\bbox{G}^{\rm V}({\bf r},{\bf r'},\omega)$ represents 
the contribution of the direct waves from the radiation sources
in an unbounded medium, which is vacuum in our case, $f$ and
$s$ denote the layers where the field point and source point
locate, $\delta_{fs}$ is the usual Kronecker symbol, and the 
scattering Green tensor 
$\bbox{G}^{(fs)}({\bf r},{\bf r'},\omega)$ describes the 
contribution of the multiple reflection ($f=s$) and
transmission ($f\neq s$) waves. 
In particular, $\bbox{G}^{(13)}({\bf r},{\bf r'},\omega)$ and 
$\bbox{G}^{(33)}({\bf r},{\bf r'},\omega)$ read as 
\cite{35}
\begin{eqnarray}
\label{B2}
\lefteqn{
       \bbox{G}^{(13)}({\bf r},{\bf r'},\omega)
       = {ik_3\over 4\pi} \sum_{e,o} 
       \sum_{n=1}^\infty \sum_{m=0}^n 
}
\nonumber\\ 
       &&\hspace{8ex}\times
       \Biggl\{ \frac{2n\!+\!1}{n(n\!+\!1)} 
       \frac{(n\!-\!m)!}{(n\!+\!m)!}  
       (2\!-\!\delta_{0m})
\nonumber \\ 
       &&\hspace{8ex}\times 
       \biggl[ A^{13}_M (\omega)
       {\bf M}^{(1)}_{{e \atop o}nm} ({\bf r},k_1)
       {\bf M}_{{e \atop o}nm} ({\bf r}',k_3)
\nonumber \\ 
       &&\hspace{8ex} + 
       A^{13}_N (\omega)
       {\bf N}^{(1)}_{{e \atop o}nm} ({\bf r},k_1)
       {\bf N}_{{e \atop o}nm} ({\bf r}',k_3) \biggr]\Biggr\} \ ,
\\
\label{B3}
\lefteqn{
       \bbox{G}^{(33)}({\bf r},{\bf r'},\omega)
       = {ik_3\over 4\pi} \sum_{e,o} 
       \sum_{n=1}^\infty \sum_{m=0}^n 
}
\nonumber\\ 
       &&\hspace{8ex}\times
       \Biggl\{ \frac{2n\!+\!1}{n(n\!+\!1)} 
       \frac{(n\!-\!m)!}{(n\!+\!m)!}  
       (2\!-\!\delta_{0m})
\nonumber \\ 
       &&\hspace{8ex}\times 
       \biggl[ C^{33}_M (\omega)
       {\bf M}_{{e \atop o}nm} ({\bf r},k_1)
       {\bf M}_{{e \atop o}nm} ({\bf r}',k_3)
\nonumber \\ 
       &&\hspace{8ex}
       + C^{33}_N (\omega)
       {\bf N}_{{e \atop o}nm} ({\bf r},k_1)
       {\bf N}_{{e \atop o}nm} ({\bf r}',k_3) \biggr]\Biggr\} \ ,
\end{eqnarray}
where
$$
       k_1=k_3={\omega\over c}, \qquad 
       k_2=\sqrt{\epsilon(\omega)}\,{\omega\over c}\, 
$$
and ${\bf M}$ and ${\bf N}$ represent TM- and TE-waves, respectively,
\begin{eqnarray} 
\label{B4}
       {\bf M}_{{e \atop o}nm}({\bf r},k) 
       &&= \mp {m \over \sin\theta} j_n(kr)
       P_n^m(\cos\theta) {\sin\choose\cos} (m\phi) {\bf e}_{\theta}
\nonumber\\
       && - j_n(kr) \frac{dP_n^m(\cos\theta)}{d\theta} 
       {\cos\choose\sin} (m\phi) {\bf e}_{\phi} \ ,
\\
\label{B5}
       {\bf N}_{{e \atop o}nm}({\bf r},k) 
       &&= {n(n+1)\over kr} j_n(kr) 
       P_n^m(\cos\theta) {\cos\choose\sin} (m\phi) {\bf e}_r
\nonumber\\
       && + {1\over kr} \frac{d[rj_n(kr)]}{dr} 
       \Biggl[
       \frac{dP_n^m(\cos\theta)}{d\theta} 
       {\cos\choose\sin} (m\phi) {\bf e}_{\theta}
\nonumber \\
       && \mp {m \over \sin\theta}
       P_n^m(\cos\theta) {\sin\choose\cos} (m\phi) {\bf e}_{\phi}
       \Biggr], 
\end{eqnarray}
with $j_n(x)$ being the spherical Bessel function of the first kind
and $P_n^m(x)$ being the associated Legendre function. 
The superscript ${(1)}$ in Eq.~(\ref{B2}) indicates that 
in Eqs.~(\ref{B4}) and (\ref{B5}) the spherical Bessel function $j_n(x)$ 
has to be replaced by the first-type spherical Hankel function $h^{(1)}_n(x)$.

The coefficients $A^{13}_{M,N}$ and $C^{33}_{M,N}$ in Eqs.~(\ref{B2}) 
and (\ref{B3}) are defined by
\begin{eqnarray}
\label{B6}
       A^{13}_{M,N} (\omega)
       &&= \frac { T_{F1}^{M,N} T_{F2}^{M,N} T_{P1}^{M,N} }
       { T_{P1}^{M,N} + T_{F1}^{M,N} R_{P1}^{M,N} R_{F2}^{M,N} }\ ,
\\
\label{B9}
       C^{33}_{M,N} (\omega)
       && = { A^{13}_{M,N} \over T_{P2}^{M,N}}
       \left[ {R_{P2}^{M,N} \over T_{F1}^{M,N}}  
       +      {R_{P1}^{M,N} \over T_{P1}^{M,N}}\right] ,
\end{eqnarray}
where
\begin{eqnarray}
\label{B10}
       R^M_{Pf} &&= \frac 
       { k_{f+1} H'_{(f+1)f}H_{ff} - k_f H'_{ff}H_{(f+1)f} }
       { k_{f+1} J_{ff}H'_{(f+1)f} - k_f J'_{ff}H_{(f+1)f} } ,
\\
\label{B11}
       R^M_{Ff} &&= \frac 
       { k_{f+1} J'_{(f+1)f} J_{ff} - k_f J'_{ff}J_{(f+1)f} }
       { k_{f+1} J'_{(f+1)f} H_{ff} - k_f J_{(f+1)f} H'_{ff} },
\\
\label{B12}
       R^N_{Pf} &&= \frac 
       { k_{f+1} H_{(f+1)f} H'_{ff} - k_f H_{ff}H'_{(f+1)f} }
       { k_{f+1} J'_{ff} H_{(f+1)f} - k_f J_{ff}H'_{(f+1)f} },
\\
\label{B13}
       R^N_{Ff} &&= \frac 
       { k_{f+1} J_{(f+1)f} J'_{ff} - k_f J_{ff}J'_{(f+1)f} }
       { k_{f+1} J_{(f+1)f} H'_{ff} - k_f J'_{(f+1)f}H_{ff} },
\\
\label{B14}
       T^M_{Pf} &&= \frac 
       { k_{f+1} \left( J_{(f+1)f}H'_{(f+1)f} 
       - J'_{(f+1)f}H_{(f+1)f} \right) }
       { k_{f+1} J_{ff}H'_{(f+1)f} - k_f J'_{ff}H_{(f+1)f} },
\\
\label{B15}
       T^M_{Ff} &&= \frac 
       { k_{f+1}\left(J'_{(f+1)f} H_{(f+1)f} 
       - J_{(f+1)f} H'_{(f+1)f} \right) }   
       { k_{f+1} J'_{(f+1)f} H_{ff} - k_f J_{(f+1)f} H'_{ff} },
\\
\label{B16}
       T^N_{Pf} &&= \frac 
       { k_{f+1} \left( J'_{(f+1)f} H_{(f+1)f} 
       - J_{(f+1)f}H'_{(f+1)f} \right) }
       { k_{f+1} J'_{ff} H_{(f+1)f} - k_f J_{ff}H'_{(f+1)f} },
\\
\label{B17}
       T^N_{Ff} &&= \frac 
       { k_{f+1}\left( J_{(f+1)f} H'_{(f+1)f} 
       - J'_{(f+1)f}H_{(f+1)f} \right) }
       { k_{f+1} J_{(f+1)f} H'_{ff} - k_f J'_{(f+1)f}H_{ff} }
\end{eqnarray}
with
\begin{eqnarray}
\label{B18}
       J_{il} &&= j_n(k_iR_l) ,
\\
\label{B19}
       H_{il} &&= h^{(1)}_n(k_iR_l) ,
\\
\label{B20}
       J'_{il} &&= {1\over \rho} 
       { d[\rho j_n(\rho)] \over d\rho } 
       \bigg|_{\rho=k_iR_l} ,
\\
\label{B21}
       H'_{il} &&= {1\over \rho} 
       { d[\rho h^{(1)}_n(\rho)] \over d\rho } 
       \bigg|_{\rho=k_iR_l}.
\end{eqnarray}
Note that $A^{13}_{M,N}$ and $C^{33}_{M,N}$ are functions of $n$
but not of $m$. When the atom is positioned at the cavity center, 
we have \cite{9}
\begin{eqnarray}
\label{B22}
       {\bf M}_{{e \atop o}nm} ({\bf r},k) \Big|_{kr \rightarrow 0}
       \longrightarrow (kr)^n ,
\\
\label{B23}
       {\bf N}_{{e \atop o}nm} ({\bf r},k) \Big|_{kr \rightarrow 0}
       \longrightarrow (kr)^{n-1}.
\end{eqnarray}
In this case, only TM-waves with $n$ $\!=$ $\!1$ contribute and the 
Eq.~(\ref{B3}) simplifies to 
\begin{equation}
\label{B24}
     \bbox{G}^{\rm R} \equiv \bbox{G}^{(33)}({\bf r},{\bf r'},\omega)
       \Big|_{r=r' \rightarrow 0}
       = {i\omega\over 6\pi c} C^{33}_N(\omega) \bbox{I} .
\end{equation}
Similarly, Eq.~(\ref{B2}) reduces to
\begin{eqnarray}
\label{B25}
\lefteqn{
      G^{(13)}_{rz}({\bf r},{\bf r'},\omega) 
      \Big|_{r'\rightarrow 0} = 
      {i\cos\theta\over 2\pi r} h_1^{(1)}(k_3 r) 
      A^{13}_N(\omega),
}
\\ &&
      G^{(13)}_{\theta z}({\bf r},{\bf r'},\omega) 
      \Big|_{r'\rightarrow 0} = 
      - {i\sin\theta\over 4\pi r} {d[rh_1^{(1)}(k_3 r)]\over dr} 
      A^{13}_N(\omega) ,
\nonumber \\ 
\label{B26} &&\hspace{30ex} 
\\ 
\label{B27} &&
      G^{(13)}_{\phi z}({\bf r},{\bf r'},\omega) 
      \Big|_{r'\rightarrow 0} = 0 .
\end{eqnarray}


\section{Derivation of Eq.~(\protect\ref{E41C})}
\label{appC}

For an atom at the cavity center and $z$-oriented dipole,  
from Eqs. (\ref{e26a}) and (\ref{B25}) -- (\ref{B27}) 
we derive  
\begin{equation}
\label{C1}
        F_r({\bf r},{\bf r}_{\rm A},\omega_{\rm A}) =
        {\cal O}(\rho^{-2}) ,
\end{equation}
\begin{eqnarray}
\lefteqn{
        F_\theta({\bf r},{\bf r}_{\rm A},\omega_{\rm A}) = 
        -{k_{\rm A}^2\mu\sin\theta \over 4\pi^2\epsilon_0\rho}
}
\nonumber\\ 
\label{C2} &&\hspace{2ex} \times   
        \int_0^\infty \!\! d\omega\ {\rm Im}\! 
        \left[A^{13}_N(\omega) e^{i\omega\rho/c} \right]
        \zeta(\omega_{\rm A}\!-\!\omega)
        + {\cal O}(\rho^{-2}) ,
\end{eqnarray}
\begin{equation}        
\label{C3} 
        F_\phi({\bf r},{\bf r}_{\rm A},\omega_{\rm A}) = 0 .
\end{equation}
Recalling the relation $\zeta(x)$ $\!=$ $\!i/(x+i0)$,
we perform the $\omega$-integration in Eq.~(\ref{C2}) to obtain
\begin{eqnarray}
\label{C4}
\lefteqn{
        \int_0^\infty d\omega\, {\rm Im}\! 
        \left[A^{13}_N(\omega) e^{i\omega\rho/c} \right]
        \zeta(\omega_{\rm A}-\omega)
}
\nonumber\\ &&\hspace{5ex}
        = -{1\over 2} 
        \int_0^\infty d\omega \,\frac 
        {\left\{ A^{13}_N(\omega) e^{i\omega\rho/c} -{\rm c.c.}
        \right\} }
        {\omega-(\omega_{\rm A}+i0)}
\nonumber\\ &&\hspace{5ex}
        \simeq -i\pi A^{13}_N(\omega_{\rm A}) 
        e^{i\omega_{\rm A}\rho/c} ,
\end{eqnarray}
where we have (approximately) replaced $A^{13}_N(\omega)$ by 
$A^{13}_N(\omega_{\rm A})$, extended the lower limit of 
the integral to $-\infty$, and applied contour-integration 
techniques. Combining Eqs.~(\ref{e26}), (\ref{e26a}), (\ref{e41b}),
(\ref{C1}), (\ref{C3}), and (\ref{C4}), it is not difficult to
prove that 
\begin{equation}
\label{C5}
        W \simeq \hbar\omega_{\rm A}\,
        \frac{ |A^{13}_N(\omega_{\rm A})|^2}
        {1+{\rm Re} C^{33}_N(\omega_{\rm A}) }\,.
\end{equation}
Taking into account that the
free-space value $W_0$ $\!=$ $\!\hbar\omega_{\rm A}$ is obtained by setting 
\mbox{$A^{13}_N$ $\!=$ $\!1$} and $C^{33}_N$ $\!=$ $\!0$, Eq.~(\ref{C5})
just yields Eq.~(\ref{E41C}).


\end{document}